\documentclass[letterpaper,10pt]{article}
\usepackage[utf8]{inputenc} 
\usepackage{url}
\usepackage{graphicx}
\usepackage{amsmath,amssymb}
\usepackage{booktabs}
\usepackage{wasysym}
\usepackage{xcolor}
\usepackage[numbers]{natbib}

\usepackage[affil-it]{authblk}
\usepackage{hyperref}

\title{Information Overload in Group Communication\\
\large From Conversation to Cacophony in the Twitch Chat}

\author[1]{Azadeh Nematzadeh\thanks{Corresponding author: azadeh.n@gmail.com}}
\author[2]{Giovanni Luca Ciampaglia}
\author[1]{Yong-Yeol Ahn}
\author[1,2]{Alessandro Flammini}

\affil[1]{Center for Complex Networks and Systems Research, Indiana University, Bloomington, USA}
\affil[2]{Network Science Institute, Indiana University, Bloomington, USA}

\date{}

\begin{document}

\maketitle

\begin{abstract}
Online communication channels, especially social web platforms, are rapidly replacing traditional ones. Online platforms allow users to overcome physical barriers, enabling worldwide participation. However, the power of online communication bears an important negative consequence --- we are exposed to too much information to process. Too many participants, for example, can turn online public spaces into noisy, overcrowded fora where no meaningful conversation can be held. 
Here we analyze a large dataset of public chat logs from Twitch, a popular video streaming platform, in order to examine how information overload affects online group communication. We measure structural and textual features of conversations such as user output, interaction, and information content per message across a wide range of information loads. 
Our analysis reveals the existence of a transition from a conversational state to a cacophony --- a state of overload with lower user participation, more copy-pasted messages, and less information per message. 
These results hold both on average and at the individual level for the majority of users.
This study provides a quantitative basis for further studies of the social effects of information overload, and may guide the design of more resilient online communication systems.  
\end{abstract}


\section*{Introduction}
The rapid growth of the social web and the penetration of mobile devices are placing people under a constant barrage of text messages, videos, and sound. Online social communication channels such as social media or other online communities allow people to receive virtually unlimited information from a gamut of sources. Yet, the information processing capacity of humans is limited. A classic example is the size of the working memory, which is limited to about seven discrete items~\cite{miller1956magical}. The speed of reading and mentally processing written texts is also limited by the physical constraints of our eyes and brain functions~\cite{Rayner1986,Duncan1994}. More fundamentally, the number of social ties an individual can sustain is limited both in the offline~\cite{Dunbar2002} and online world~\cite{Goncalves2011}.

When the amount of information exceeds these cognitive limits, \emph{information overload} may ensue. Intuitively, this can be  defined as the state in which one cannot make sense of incoming stimuli. While there is ample evidence of information overload in individual settings~\cite{eppler2004concept,Janssen2006,Savolainen2007,Bawden2009,anderson2012competition,case2012looking}, its \emph{social} implications, i.e. the effects on collective communication, are still largely unknown. Only recently, some research has started looking into the effect of information overload into social contagion and email communication~\cite{hodas2012visibility,gomez2014quantifying,kooti2015evolution}.

Group communication is an excellent setting where collective outcomes may be impaired by information overload, and new online communication channels provide ideal experimental platforms for analyzing this phenomenon.

Online communication technologies create frameworks for individuals to troubleshoot problems, discuss questions, or leisurely chat with each other. These frameworks create a \emph{discourse environment}, or \emph{virtual public}, that introduces a new form of communication in which a potentially limitless number of people can join a single conversation. Consequently, such a novel form of mass interaction~\cite{whittaker2003dynamics} may result in information overload among social media users~\cite{eppler2004concept,gomez2014quantifying}.
 
Evidence of information overload on online group communication has been previously reported for the case of Usenet newsgroups~\cite{jones2004information} and IRC (Internet Relay Chat) channels~\cite{jones2008empirical}. As the number of messages increases, participants cannot follow the conversation and their chances of replying to earlier messages decrease. Thus, even though the total number of messages increases, the number of messages per user may decrease. Assuming that users will tend to avoid this kind of situations, it was predicted that the overall group size would be limited. Indeed, it has been found that IRC rooms typically saturate at up to 300 users, with a maximum of about 40 participants actively talking~\cite{jones2008empirical}. Similar observations hold for Usenet~\cite{jones2004information}; recent ethnographic studies are in support of these conclusions~\cite{hamilton2014streaming}.

While these studies provide evidence that information overload has macroscopic effects on group communication, the scale of the data obtained from legacy communication systems such as IRC may not be large enough to draw strong statistical claims. The activity rates recorded on those platforms are too low compared to what large crowds can produce on modern social media platforms. Moreover, it is not clear how overload affects the structure and content of communication.

\begin{figure}[t!]
\centering
\includegraphics[width=1\columnwidth]{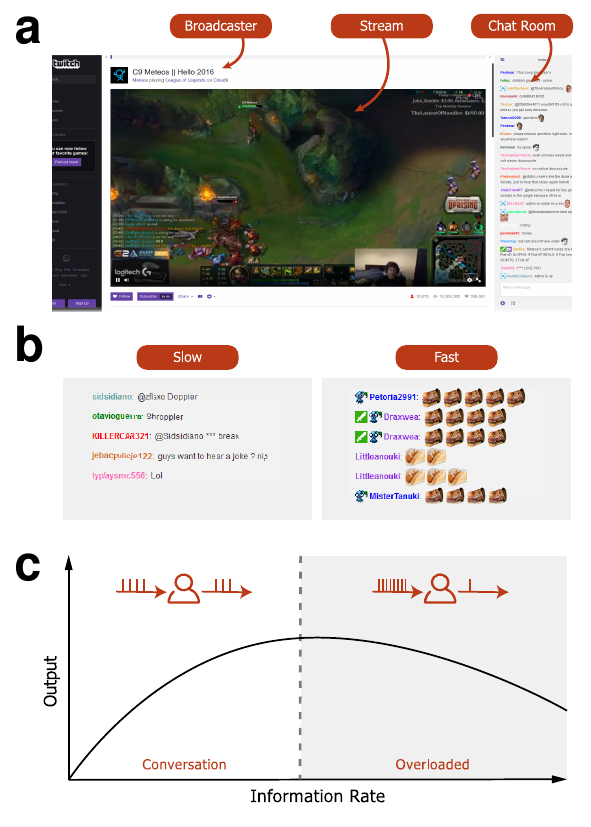}
\caption{Study settings and illustration of our hypothesis. (a) user interface of a Twitch stream. (b) excerpts of chat logs from two chat rooms, corresponding to a setting with low rate of production of messages (left) and a high one (right), respectively. (c) we hypothesize that information overload gradually impairs group communication: as the rate of information processed by individual users grows, the output of each participant, as measured, for example, by the number of written messages, increases (left). Moreover, messages carry the information content of actual discourse. As the rate grows past a threshold, output decreases (right). The information content of messages decreases as well, as exemplified by repetition (e.g. \emph{copy-pasting}) and by the disproportionate use of non-verbal symbols (e.g. emojis).}
\label{fig:settings}
\end{figure}

\begin{figure}
\centering
\includegraphics[width=\columnwidth]{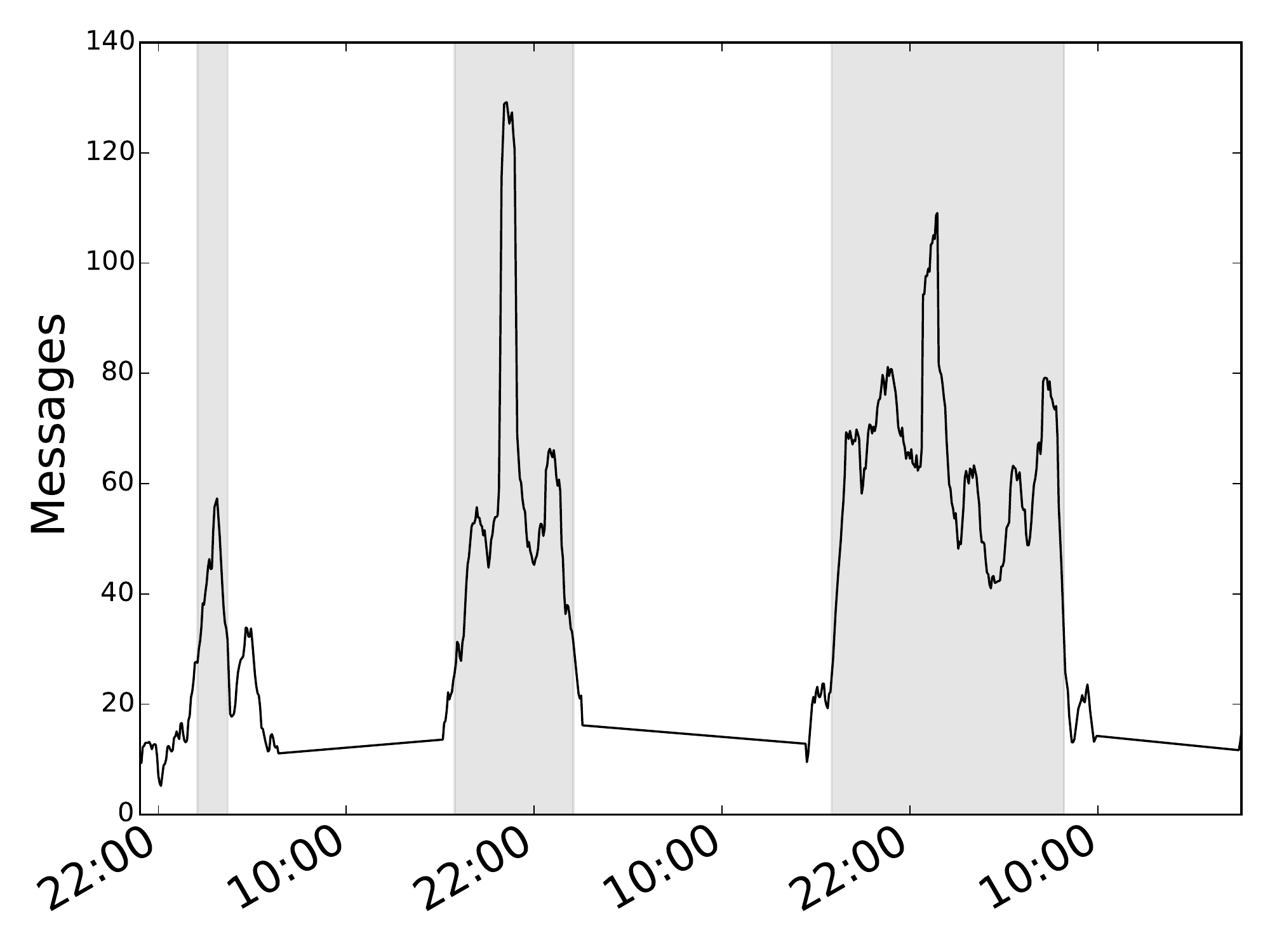}
\caption{A typical time series of the volume of chat messages in a room. We sampled the number of messages posted in the room every 5 minutes. Here peak activity corresponds to roughly one message every 3.8s. We infer the periods when a live broadcast took place (shaded areas) from sudden shifts in activity (see Methods).}
\label{fig:broadcasts}
\end{figure}

Overcoming these limitations, here we analyze a large dataset of chat logs from Twitch\footnote{\url{http://www.twitch.tv/}.}, a popular video sharing and streaming platform. On Twitch, people can broadcast in real time a \emph{stream} --- usually a video feed of their screen --- to other users, and share videos of past broadcasts, see Figure~\ref{fig:settings}(a).
As in many video-sharing platforms, like the popular YouTube, users can leave comments to any stream but, unlike those, on Twitch, users write messages into an \emph{interactive and real-time} chat room displayed prominently on the side of the stream, see Figure~\ref{fig:settings}(b).
As a result, viewers are exposed to a live flow of information, as in a traditional IRC channel. Unlike traditional IRC, however, popular Twitch streams are often watched by massive audiences ranging in the hundreds of thousands of viewers~\cite{kaytoue2012watch}, resulting in unprecedented rates of message production. For example, in Figure~\ref{fig:broadcasts}, we show the rate at which messages are posted in the chat window of a moderately active channel in our dataset, over a 60-hour period. The shaded areas, which roughly correspond to the time of live broadcasts, see peaks of more than 100 messages every 5 minutes (approximately 1 message every 3 seconds).

Aside from the top popular channel, the majority of streams that feature any content at all are watched only by few users\footnote{The majority of streams do not feature any content, and are not watched by anybody. Because we have only access to the logs of chat messages, we do not have any data about these streams.}. Moreover, broadcasters typically stream live only for a few hours a day, and for the rest of the time the stream is inactive. 
During these periods of inactivity those viewers who connect to the stream can choose to watch recordings of past broadcasts.

Even when focusing on a single stream, there is a great deal of variation in messaging activity; for the most successful streamers this activity typically spans several orders of magnitude, depending on the time of the day and on whether a game, or other type of performance, is being streamed live or not.
This heterogeneity allows us to examine user behavior across a wide range of conditions --- from times when messages are posted very slowly and by a handful of people, to times when a huge volume of messages are pounding the chat window.

Our main hypothesis is that the \emph{output} of users depends on the rate at which they receive information, and that this identifies two distinct phases, which we call \emph{conversation} and \emph{overload}, respectively; see Figure~\ref{fig:settings}(c). In the conversation phase, as the rate of incoming information increases, user output increases, since more incoming messages elicit more replies. 
We define the overload phase as the case in which more information corresponds to a decrease in the output. 
Moreover, as information overload impairs basic cognitive capabilities, we expect that users will resort to simpler, and more stereotyped utterances, or that they will simply repeat what others are writing, as in a chorus, or a cacophony. To summarize, we are interested in the following research questions:
\\[.5em]
\textsc{rq}~1. \emph{Are there two phases --- conversation and overload --- in the Twitch data?}\\[.5em]
\textsc{rq}~2. \emph{Is the overload phase marked by a decrease of participation (output), as measured by the average number of messages per user, in the Twitch chat data?}\\[.5em]
\textsc{rq}~3. \emph{Does the overload phase correspond to a decrease of the information content (i.e. the number of bits needed to encode a message) produced by each individual message? Are there any visible changes to the nature of conversations?}\\[.5em]
\textsc{rq}~4. \emph{Does the transition from the conversation to the overload phase happen abruptly, or is there instead a gradual deterioration?}\\[.5em]
\textsc{rq}~5. \emph{Can we quantify information overload at the individual level?}

\subsection*{Ethics Statement}

This research was performed in compliance with the IRB regulations of Indiana University (protocol no. 1410552242). 
Since the identity of users on Twitch (i.e. the user name) is not relevant to our analysis and may pose unnecessary risk to the users, we first anonymized all data using a non-cryptographic hash function\footnote{The code of the non-cryptographic hash function used to anonymize the data is available at \url{https://github.com/aappleby/smhasher}.}; all results presented in the paper are obtained from this anonymyzed dataset. 

\section*{Related work}
In this section we give a brief survey of relevant literature on attention, information overload, and collective phenomena. Because we test our research questions using data from Twitch, and because part of the terminology used in the paper draws directly from the lexicon typically used within its community, we provide also a brief survey of research on Twitch and video-sharing communities.

\subsection*{Information Overload}

Information overload has been studied and discovered independently in several fields of investigation. In an information-rich world, allocating attention among different activities or stimuli becomes a problem~\cite{Simon1971}. This is well known to social scientists, as individuals must distribute time among interpersonal ties in more or less even manner~\cite{backstrom2011center}, or drop ties altogether~\cite{Miritello2013}.

As the number of individuals or messages in a virtual public increases, the amount of communication expands to a degree that may not be conceivable by group participants. Thus, individuals may not be able to follow a one-to-one, one-to-many, or many-to-many conversation. As a result of participants experiencing information overload, structure and dynamics of the discourse can be highly affected and reshaped~\cite{jones2002boundaries,jones2000virtual}. For example, individuals may only reply to certain topics, reply in a shorter and simpler manner, or may stop participating altogether~\cite{hiltz1985structuring,rafaeli1997networked}.
The size of the virtual public and the volume of communication can also influence the chances of interacting. When the volume of information is high either in the form of long or complex messages, or in the form of a large number of messages, the chance of taking part in group communication is low~\cite{jones2004information}.

Information overload is also a topic in cognitive psychology, organization science, and marketing~\cite{chaiken1987heuristic,eppler2004concept}. It can cause ``attentional conflict'' if individuals cannot focus on selecting one option out of many. This form of cognitive overload can decrease the performance of individuals~\cite{baron1986distraction}. In marketing, consumers are assumed to be generally overloaded~\cite{Keller1987}, and companies have the chances to win their attention by means of advertising~\cite{anderson2012competition}. In online social networks, user-generated memes compete for limited attention~\cite{weng2012competition}, but repeated exposure to a given piece of information improve the chances of adoption~\cite{feng2014competing}.

Addressing the problem of information overload lies at the heart of information retrieval, search, and data mining~\cite{belkin1992information,baeza1999modern}. Indeed, search engines such as Google were initially built with the objective of organizing the world's information.
Improvement in the design of systems and interfaces can mitigate information overload by omitting information in excess~\cite{tufte1983visual,Maes1994,chan2001use}. Social media such as Facebook do indeed resort to algorithmic filtering and ranking to solve the problem of overload~\cite{Bakshy2015}. This may lead users to be trapped into so-called \emph{filter bubbles}, information environments with low content diversity and high social reinforcement~\cite{Nikolov2015}, though the exact origin of these (i.e., algorithmic vs social) is still subject of investigation. On the other hand, models show that the notion of trust can help overcome overload in recommendation systems~\cite{Walter2008}.

\subsection*{Twitch}

Twitch started in 2007 as a ``social TV'' experiment, under the name of justin.tv. Users could broadcast video streams into their own channels. With the rise of popularity of electronic sports, or \emph{eSports}, it has rapidly become one of the most popular live streaming platforms on the Internet~\cite{Blackburn2014,Kwak2015}. The majority of streams on Twitch are in fact about video games, played by either amateur or professional players~\cite{hamilton2014streaming}.
According to in-house statistics, Twitch rivals with traditional television networks (e.g. CNN) in terms of viewership, with 100M monthly viewers in 2015. It also accounts for 1.8\% of the overall peak-time Internet traffic, ranking behind only to Netflix, Google, and Apple~\cite{twitch2015traffic}. These sheer numbers give an idea of the size of the Twitch phenomenon and of the worthiness of its conversation dynamics for researchers interested in information overload and related phenomena.

\section*{Methods}

\begin{figure}[t]
\centering
\includegraphics[width=\columnwidth]{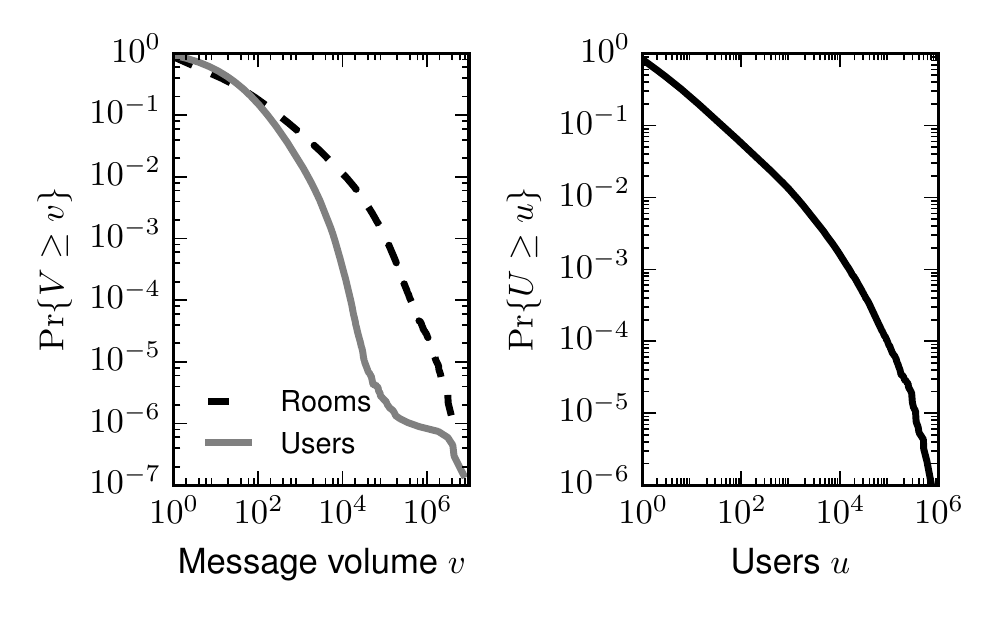}
\caption{Distributions of messaging activity and stream popularity. Left: distribution of the number of messages per room (black dashed line) and per user (solid grey line). Right: distribution of the number of users per channel. All quantities show broad distributions.}
\label{fig:ccdf}
\end{figure}

Our data include all messages posted to any public Twitch chat room within the period between August 26 and November 10, 2014 (76 days). We counted a total number of 1,275,396,751 messages, posted in 927,247 channels (average: 1,375 messages per channel) by 6,716,014 users (average: 190 messages per user).
Of all channels, 532,094 were active for at least two days, 319,451 had only one active user, and 166,870 had more than 100 messages.
Of all users, 4,930,052 were active for at least two days, 5,015,079 participated in at least two channels, and 1,032,766 posted at least 100 messages.
In these data, a user (resp. channel) is logged only if they sent (received) a message. Therefore users (or channels) who were inactive during the observation window are not included in our data, and so it is important to understand that the above figures may not reflect the total activity of \emph{registered} accounts --- viewers or broadcasters --- on Twitch.

The above quantities are, as one would expect, highly heterogeneous.
The number of messages produced in a channel (Figure~\ref{fig:ccdf} left) over the course of the whole observation period spans several orders of magnitude, with a median of 7 messages.
The same happens for channel popularity (Figure~\ref{fig:ccdf} right) and total messaging activity of users (Figure~\ref{fig:ccdf} left), some of which have written more than one million messages, while most only a few.

Besides the number of produced messages --- the ``output'' in Figure~\ref{fig:settings}(c) --- we compute several content-based metrics to detect the onset of overload, which we describe later in this section. However, the nature of our data poses several challenges to this task.

First of all, the data do not record the starting and ending of broadcasts. Broadcasters typically stream only for a few hours a day, but the chat room is available without interruption. As a result, messaging activity in a channel exhibits over the course of the day strong, brief surges, interspersed between long periods of silence, as shown in Figure~\ref{fig:broadcasts}. Since conversation and overload are more likely to occur during broadcasts, we filter out periods of inactivity, using a simpler clustering heuristic.

The second challenge has to do with the presence of messages generated by non-human accounts, or \emph{bots}. Indeed, it is common for broadcasters to automate the management of their channels, and a preliminary manual inspection also revealed the abundance of bot-generated messages.
There are many different types of bots operating on Twitch. Some report the status of the game (e.g. scores) or player (e.g. ranking), some post advertisement from the broadcaster (e.g. links to off-site pages), and some greet users upon logging into the channel. Some bots generate messages in several rooms, while others are active only in one.

Because bots do not really have cognitive limits, they can skew the estimation of our overload features. Thus it is important to remove them. At the same time, it is reasonable to expect that bot activity contributes, as much as humans activity, to trigger the onset of the phase of overload. Therefore, in computing the rate of processed information --- the $x$ axis in Figure~\ref{fig:settings}(c) ---  we do not remove the contributions from bots, but we remove them from the output of users --- the $y$ axis. Similarly, we filter out bot-generated messages when estimating content-based metrics.

The following two subsections give details about data cleaning procedures. We then describe the metrics used to detect the presence of overload in chat conversations.

\subsection*{Broadcast Detection}

Let us consider a channel $c$. To detect broadcast periods we sample the volume of messages $V_c(t)$, $t = n\Delta t$, $n = 0,1,2,\ldots$, at intervals of $\Delta t = 5$~minutes. We then consider the time average of the message volume $\overline V = \langle V_c(t)\rangle_t$ and define a symbolic sequence $S_c(t)$ where:
$$
S_c(t) = \begin{cases}
\mbox{\tt A} & \mbox{if } V_c(t) \ge \overline V \\
\mbox{\tt I} & \mbox{if } V_c(t) < \overline V
\end{cases}
$$

Examining $S_c(t)$ we noticed that sub-sequences \texttt{AIA} and \texttt{IAI}, i.e. below- or above-average spikes shorter than $\Delta t$, would sometimes occur within longer sequences of \texttt{A}s or \texttt{I}s, respectively. We replaced these fluctuations with \texttt{AAA}, and \texttt{III} respectively. Finally, we defined the sequence \texttt{IIA} as the beginning of a broadcast, and \texttt{AAI} as the ending, and recorded the respective timestamps.
If two consecutive broadcasts were separated by less than 60 minutes, we merged them together, assuming that no streamer would bother starting a separate broadcasts in such a short span. Figure~\ref{fig:broadcasts} shows, as grey shaded areas, the detected broadcast periods from one example room in our data.

\subsection*{Bot Detection}
 
Based on above intuition, we devise two discriminatory features (see Table~\ref{tab:metrics}, bottom): the
average inter-message time $\tau$ and the compression ratio
$$\rho = \frac{\hat S}{S}$$
that is achieved when all messages of a given user are concatenated into a single
string of size $S$ bytes, and compressed into a string of size $\hat S \le S$.
The compression ratio $0 < \rho \le 1$ quantifies the information content of the messages of a user, in a way similar to Shannon's entropy. In using this approach we are motivated by the notion of Kolmogorov complexity~\cite{li2013introduction}. To perform the compression, we use the \textsc{deflate} algorithm, as implemented in the zlib library version 1.2.6~\cite{deutsch1996zlib}.

The rationale for these features is two-fold: bots are known to produce messages at a rate higher than what is physically possible by a human; and it is reasonable to expect that these messages will be also more stereotyped and repetitive.

\begin{figure}
\centering
\includegraphics[width=\columnwidth]{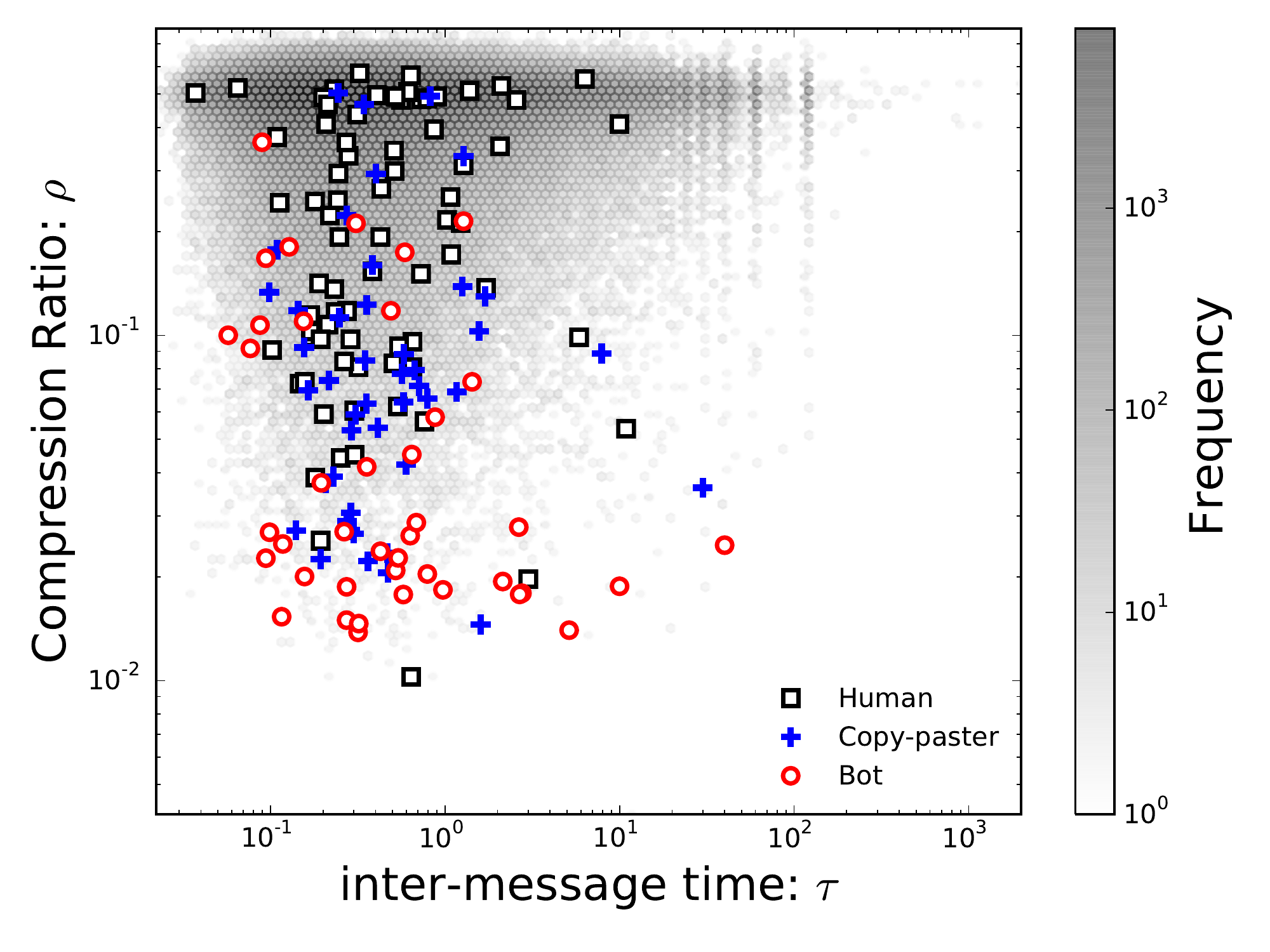}
\caption{Bot classification. Manually labeled examples for the three classes `bot', `human', and `copy-paster' are shown in the $\tau\times\rho$ feature space. As a reference, the population density of users from which these examples were drawn is shown, with shades of gray mapped into the frequency of users in each bin.}
\label{fig:bots}
\end{figure}

To detect bots, we considered all users who had been active more than one day and who had produced at least 10 messages (865,551 users), and used a stratified sampling approach on $\tau$ and $\rho$ to randomly selected from this population 256 users, see Figure~\ref{fig:bots}.
We then manually inspected all of their messages, and labeled them with one of the following categories: \emph{bot}, \emph{human}, \emph{copy-paster}, \emph{non-english}, and \emph{ambiguous}. The `copy-paster' label is meant to capture users whose complete production consists only of one or more brief, fast sequences composed by the same, copy-pasted, message. Out of 256 users, we identify 49 bots, 92 humans, and 59 copy-pasters. The remaining users were either `ambiguous' or `non-english'. We discarded these latter two groups from the following analysis.

When estimating $\tau$, we need to take into account the fact that neither bots nor humans are necessarily active at all times.
We therefore compute the average inter-message time only during `active'
periods: a user is considered in an active period if they have produced at least one message in the previous hour. If not, the inter-time is discarded and a new active session begins with the next message. The average of $\tau$ is then estimated across all sessions with at least two messages.

Figure~\ref{fig:bots} shows the distribution of labeled examples in the $\tau\times\rho$ feature space and, as a reference, the general population distribution used for sampling. Surprisingly, $\tau$ does not offer much discriminative power, while $\rho$ does.

Since the purpose of this detection task is to clean the data from bots, our objective is to detect as many bots as possible, i.e. minimize Type II errors or, equivalently, maximize the true positive rate. Therefore, we only kept users with $\rho \ge 0.44$, which corresponds to the maximum value of $\rho$ observed for a bot among our examples. The total number of users removed is thus equal to 43,026 (0.5\% of the reference population).

\begin{table}[th]
    \centering
    \footnotesize
    \begin{tabular}{@{}c l   @{\extracolsep{\fill}}}
    \toprule
    \textbf{Symbol} & \textbf{Metric} \\
    \midrule
        $V$            & Message volume           \\
        $U$            & Active users             \\
        $M_u$          & Messages per user        \\
        $l_m$          & Message length           \\
        $p_{\sf ?}$    & Questions frequency      \\
        $p_{\sf @}$    & User mentions frequency  \\
        $p_d$          & Discourse markers frequency \\
        $p_{\smiley}$  & Emoticons frequency      \\
        $\rho_c$       & Compression ratio        \\      
    \midrule
        $\rho$         & Compression ratio       \\      
        $\tau$         & Inter-message time       \\
    \bottomrule
    \end{tabular}
    \caption{Metrics of information overload (top) and bot detection (bottom).}
    \label{tab:metrics}
\end{table}

\subsection*{Metrics of Information Overload}

We use several metrics to detect the onset of information overload (see Table~\ref{tab:metrics}, top). The main measure is the average user output, as sketched in Figure~\ref{fig:settings}.
To compute it, we aggregate our data into chunks with a frequency of $\Delta t = 5$~minutes. That is, at each step $t = n\Delta t$, $n = 0, 1, 2, \ldots$, we sample the volume of messages $V(t)$ sent between $t$ and $t + \Delta t$, and the number of users $U(t)$ that produced those many messages. The output is then defined as the average number of messages per user $M_u = \frac{V}{U}$.

In doing so, we treat each room independently of each other. Given a message volume $V$, we then take, across all rooms, all chunks with exactly $V$ messages and compute
%
%
$M_u(V) = \langle M_u \rangle_V$, where the subscript $V$ indicates we are taking the average given $V$.

In addition to $M_u$, we compute several other textual and lexical features to characterize the shift from conversational state into overload. Underpinning all of them is the intuition that in the phase of overload users will resort to simpler, shorter, and more stereotyped messages. These are: the message length $l_m$; the frequency of questions $p_?$, i.e. messages ending with a question mark sign; The frequency of @-mention, the frequency at which a user addresses another user with of an @-mention $p_@$ in the message; the frequency of discourse markers~\cite{schiffrin1988discourse}, i.e. colloquial expressions such as ``oh,'' ``well,'' or ``of course''; the fraction of emoticons and emotes $p_{\smiley}$; and the average block compression ratio $\rho_c$.

Frequencies $p_?$ and $p_@$ were computed at the level of messages, while $p_d$ was computed at the level of words, breaking tokens in correspondence of white spaces, after transforming all text to lowercase.

To compute $p_{\smiley}$ we used a more sophisticate approach. Beside the popular emoticons (e.g. `\texttt{:-)}' or `\texttt{:-(}') in Twitch it is also customary to use  \emph{emotes} --- short text codes associated to small images that are rendered automatically in-line within the text. The Twitch software recognizes a list of approximately 190 standard emotes. Moreover, broadcasters can define additional emotes for their channel, which are available to viewers who pay a small monthly subscription fee.

We collect both kinds of emotes from a comprehensive online resource\footnote{The data of the Twitch emotes are available at: \url{https://twitchemotes.com/}.}. The total number of subscription emotes we found is 16,763. While large, our list of subscription emotes is of course  not complete.
To compute the probability of occurrence of emoticons/emotes, we break messages into $k$-shingles. Shingles are short substrings of varying size~\cite{leskovec2014mining}. We opted for shingling over a more common word tokenization strategy because emotes are often copied and pasted in sequence without white spaces between them.
The maximum length of emotes in our list is 24, thus, we varied the value of $k$ accordingly. For each value of $k$ we created a bag of shingles. We then merged all the bags, obtaining a total of $N$ distinct shingles. Finally, we defined $p_{\smiley} = \frac{N_{\smiley}}{N}$, where $N_{\smiley}$ was the number of shingles that matched any of the emotes in our list.
 
In similar fashion to the bot detection step,  we quantify the information content of each message block with compression. We compute the compression ratio of a chunk by concatenating all messages together. We then compute $\rho_c$ as the average ratio for chunks with the same number of messages $V$, across all rooms.

\subsection*{Testing Information Overload at the Individual Level}

We want to make sure that our results hold at the individual level, and are not just the by-product of computing group averages on a mixture of different individuals whose behavior does not depend on the volume of information they are exposed to. In practice, we want to quantify the extent to which the inverted U-shape curve of Figure~\ref{fig:settings}(c) holds in the population of Twitch users. To do so, we compute $M_u$ for each user (instead of as a group average above) and we regress it against $V$, for both the case when $V<V^*$ (sub-threshold), and when $V>V^*$ (supra-threshold), where $V^*$ is the threshold that gives the onset of overload, see Figure~\ref{fig:settings}(c). 

We compute slopes $\alpha_{\rm sub}$ and $\alpha_{\rm sup}$ of the two regression lines for both the sub- and supra-threshold cases. To compare these value across different users, before computing the regression we standardize the data. We expect to find four groups of users based on the sign of the slopes of two regions. The hypothesis of information overload is satisfied if both $\alpha_{\rm sub} > 0$ and $\alpha_{\rm sup} < 0$ hold. 

\section*{Results}

\begin{figure}
\centering
\includegraphics[width=\columnwidth]{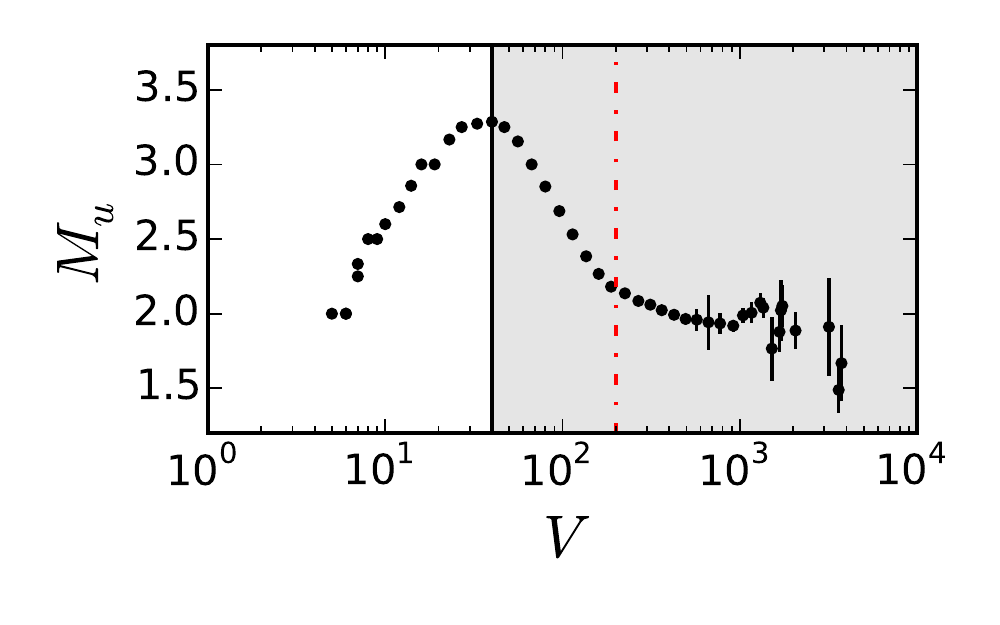}
\caption{User output (messages per user $M_u$) as a function of information rate (message volume $V$). The gray shaded area indicates the phase of overload --- for a comparison, cf.~Figure~\ref{fig:settings}(c). We include only channels with at least 1,000 messages and at least 100 user observed during the full observation window overall ($N = 43,969$). To compute $M_u$ we average across chunks of messages sampled every $\Delta t = 5\mbox{min}$. To mitigate fluctuations due to undetected bots, we use the median instead of the sample average, and consider only chunks with number of users $U > 1$. Error bars represent standard error of the mean.} \label{fig:overload}
\end{figure}

Figure~\ref{fig:overload} shows $M_u$, the average number of messages posted by a user, as a function of the message volume $V$. The plot shows an inverted U-shape: $M_u$ initially increases, peaks at the threshold $V^* \approxeq 40$ messages per 5-min block (about one message every 7.5s) and then decreases. The decay is at first abrupt, then around $V = 200$ (equivalent to one message every 0.67s) more steady.

A natural interpretation is that at low information rates an increase in activity motivates users to post more. When the activity is too high users cannot follow the flow of messages and are thus less motivated to participate, presumably because there is no expectation to meaningfully interact or even to be noticed. However, a number of alternative explanations may also be considered. 

\begin{figure}[t!]
\centering
\includegraphics[width=\columnwidth]{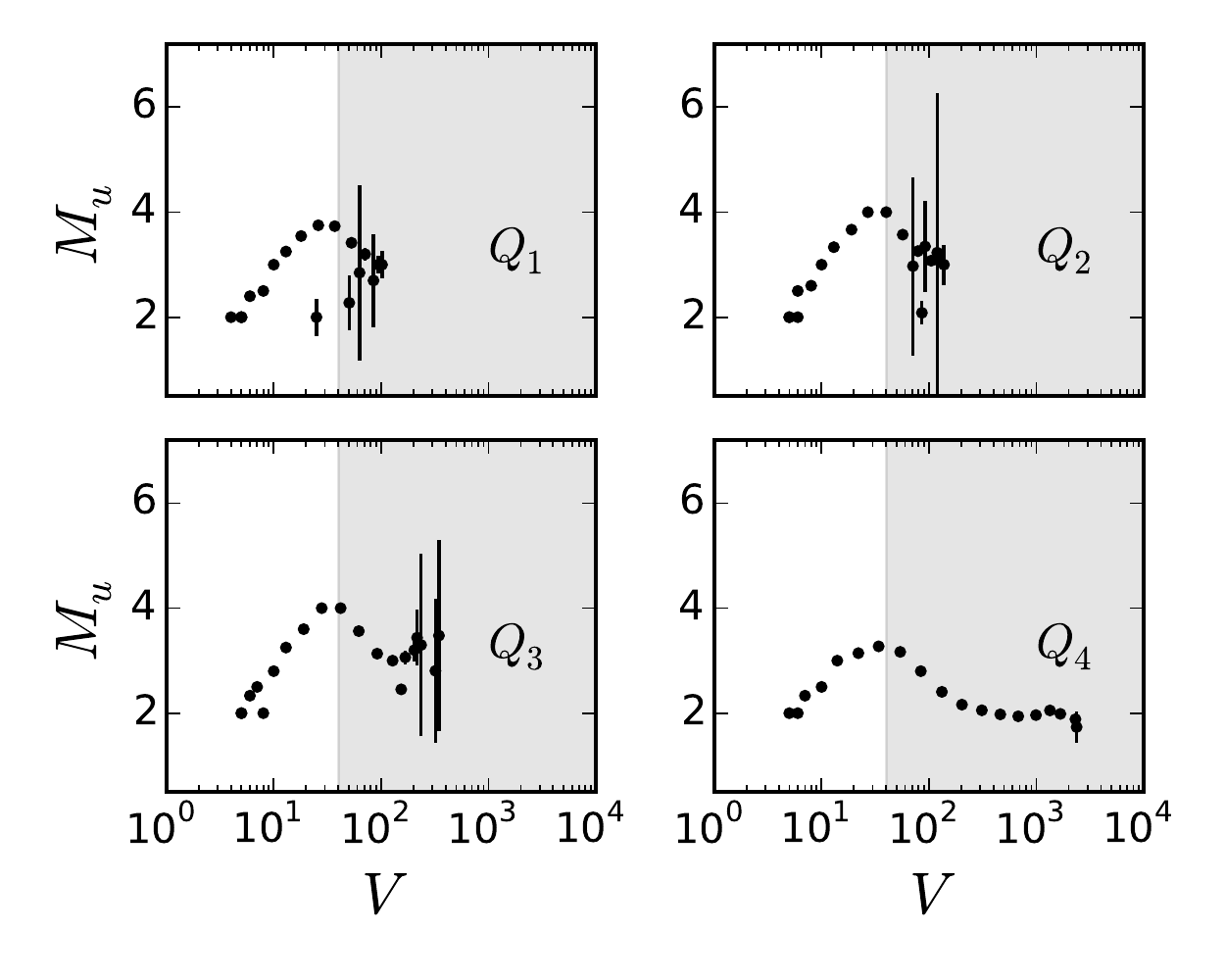}
\caption{User output (messages per user $M_u$) as a function of information rate (message volume $V$), broken down for the quartiles of the distribution of users per channels (see Figure~\ref{fig:ccdf}). The group with least active rooms is denoted by $Q_1$. To estimate $M_u$, similar steps as in Figure~\ref{fig:overload} were used, with the exception that we only consider chunks with $U > 2$.}
\label{fig:overload_quartile}
\end{figure}

The first is that the above result could be a mere artifact of aggregating together rooms with dramatically different activity levels. Low-activity rooms could be responsible for the initial increase, while high-activity rooms for the subsequent decrease. To rule out such an explanation we repeat the same exercise breaking down the population of rooms on the quartiles of the distribution of $U$, which is shown in the left panel of Figure~\ref{fig:ccdf}. Despite increased fluctuations due to smaller sample size, Figure~\ref{fig:overload_quartile} shows that the pattern still holds even when we restrict to rooms with small or intermediate activity levels.

Another alternative explanation is that the decrease in $M_u$ could be due to the fact that the overall number of users (the denominator $U$ of $M_u$) might increase while the overall number of messages $V$ stays constant. One could imagine that an increase in $U$ might happen for reasons unrelated to overload; for example, a sudden influx of users, and so our measurement might not support the hypothesis of overload. 
Note, however, that in our data $V$ and $U$ are not completely independent from one another, since our data only include users who wrote messages, i.e. we do not know the actual number of viewers of the stream. This implies that in our data an increase in $U$ must \emph{by definition} correspond to an increase in $V$, and so the above situation ($U$ increase but $V$ remains constant) cannot happen. 

The ratio $\frac{V}{U}$ could decrease also if $V$ does grow, but does so not \emph{as fast} as $U$ does. Our data could in principle support this alternative explanation. We rule it out by characterizing the joint distribution of the sub- and supra-threshold growth rate of $\frac{V}{U}$ \emph{at the individual level}; we report these results in this section below.

\begin{figure*}[t]
\centering
\includegraphics[width=\textwidth]{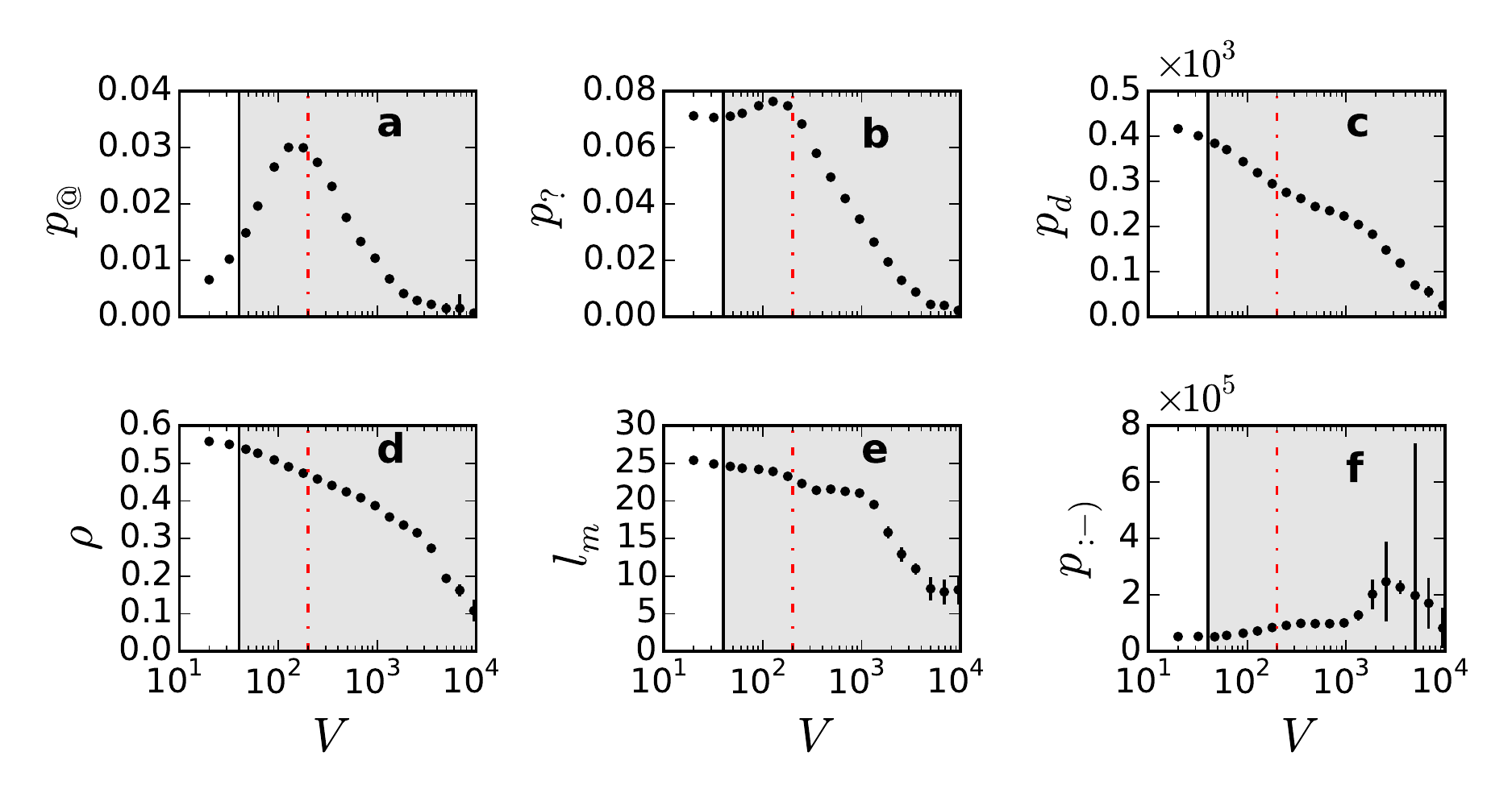}
\caption{Information-based and textual metrics of information overload, see Table~\ref{tab:metrics}. Quantities were estimated over chunks of messages collected every $\Delta t=5\mbox{min}$. The same number of rooms used for Figure~\ref{fig:overload} was used. The gray shaded area corresponds to the same region in Figure~\ref{fig:overload}, while the red dot-dashed line corresponds to $V = 200$ messages. Error bars represent the standard error of the mean.}
\label{fig:features}
\end{figure*}

Figure~\ref{fig:features} shows the results for the other metrics of overload. Our interpretation is further supported by looking at the frequency of mentions with the `@' symbol which, like in Twitter, in Twitch is used both to mention and to address other users. Figure~\ref{fig:features}(a) shows a qualitative behavior similar to that of message-per-user --- though with a later peak --- and is amenable to similar interpretation.

The frequency of questions $p_?$ (Figure~\ref{fig:features}(b)) remains approximately constant for the whole conversation phase and part of the overload phase and drops dramatically around $V = 200$. Aside from peaking earlier than $p_?$, the frequency of discourse markers $p_d$ shows a qualitatively similar behavior.

The compression ratio shown in Figure~\ref{fig:features}(d) is uniformly decreasing. As the overall activity increases the content becomes more repetitive, both within messages (e.g. emoticons repeated several times) and across messages, possibly due to increased use of cut-and-paste. The pressure on users due to overload is visible in panel (e) of Figure~\ref{fig:features}, that shows how messages get increasingly shorter as activity increases. At the same time, users resort to more emoticons and emotes (Figure~\ref{fig:features}(f)). Interestingly, in the conversation phase $p_{\smiley}$ actually \emph{decreases}, reaching the nadir approximately at the onset of the overload phase (as predicted by $M_u$).

\begin{figure}[tb]
\centering
\includegraphics[width=\columnwidth]{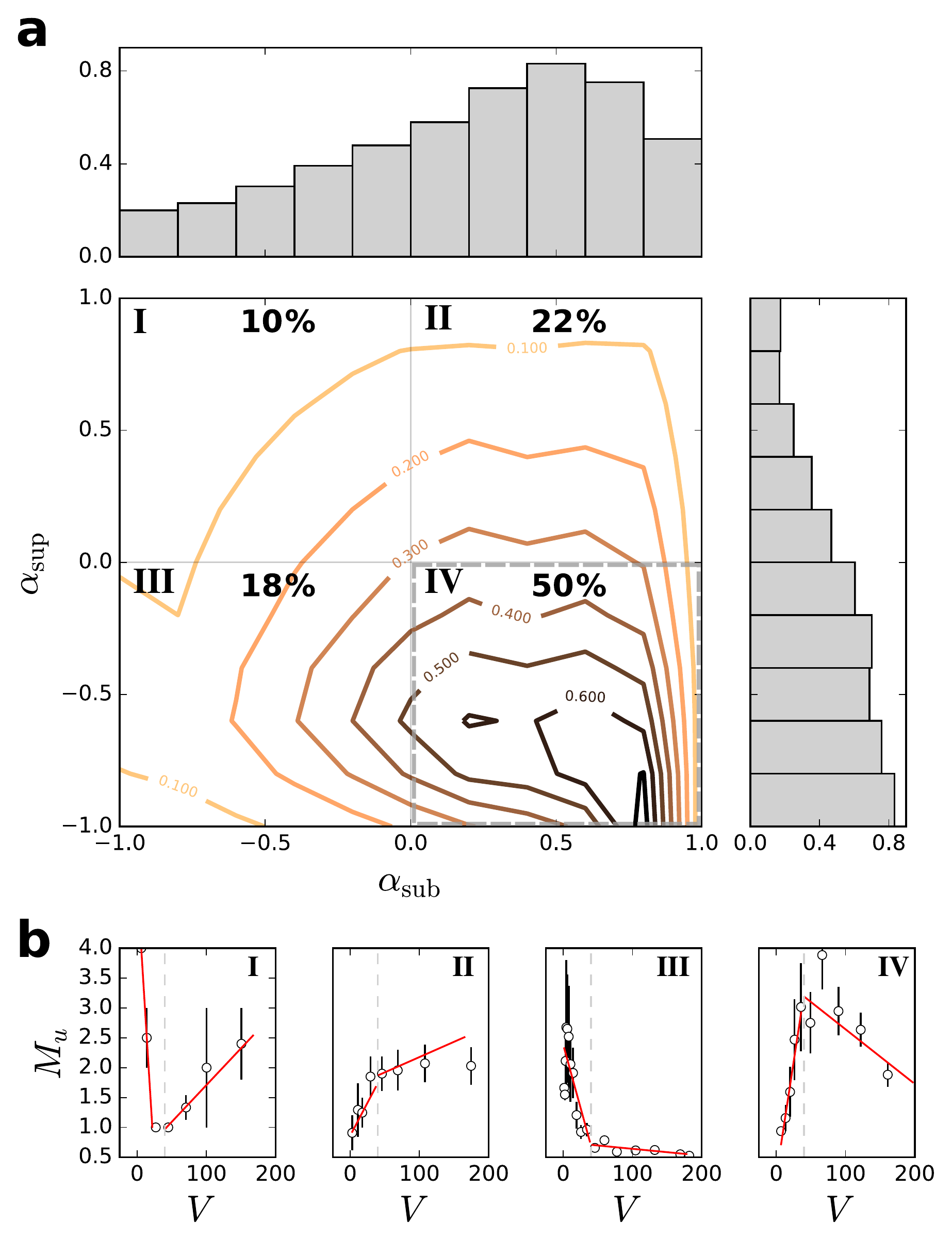}
\caption{Information overload at the individual level. (\textbf{a}) The plot shows the distribution of a sample of users in the $\alpha_{\rm sub} \times \alpha_{\rm sup}$ space for  $V^*=40$. Each point is obtained by regressing user output $M_u$ on input rate $V$ in the sub- and supra-threshold regions, respectively. Users with at least 10 observations in either regions were selected. Points that fall within the S-E quadrant (highlighted) correspond to users whose behavior is consistent with information overload. (\textbf{b}) We selected one user at random from each of the four quadrants (I-IV); in the lower panel we show their individual response curves and estimated $\alpha_{\rm sub}$ and $\alpha_{\rm sup}$ slopes (red lines). The dashed line corresponds to $V=40$.}
\label{fig:slope}
\end{figure}

Finally, to test our information overload hypothesis at the individual level, in Figure~\ref{fig:slope} we show the distribution of users in the $\alpha_{\rm sub} \times \alpha_{\rm sup}$ space for  $V^*=40$. The coefficient $\alpha_{\rm sup}$ was estimated using data in the range $V^*<V<200$, since for $V>200$ estimates of $M_u$ tend to be noisier.
Contour lines show that the majority (50\%) of users have a behavior consistent with the inverted U-shape curve model of information overload. 

\section*{Discussion}

Information overload has serious consequences on the well-being of people affected by it and it can also affect the way collective processes come about. In the case of Twitch chat conversations, we find strong evidence of overload (\textsc{rq}~1). The amount of information users are exposed to, which here we operationalize as the volume of messages $V$ that flow through the chat, seems to dictate a distinct drop in participation, which we measure as the average number of messages per user $M_u$ (\textsc{rq}~2). This is consistent with previous accounts of overload, e.g.~\cite{kooti2015evolution}.

The transition seems to occur between two states, with strong visible changes between the two (\textsc{rq}~3). The first state is marked by several characteristics akin to conversation: messages are long, varied, and marked by a typical conversational tone; users interact with each other by means of questions and direct mentions; the usage of emoticons is kept to a minimum. The second state is radically different: messages become shorter and more stereotypical, as evidenced by their higher compressibility; conversational markers disappear, replaced by an increase of emoticons; users stop interacting with each other. Such a phase is more akin to a \emph{cacophony} --- a discordant mixture of distinct voices overlapping each other.

Regarding the nature of this transition (\textsc{rq}~4), we find that this is not an abrupt process, but rather a gradual one, marked by two distinct turning points. The first corresponds to the peak of user participation, around $V = 40$ messages every $\Delta t = 5\mbox{min}$.
The second one, located around $V = 200$ messages per $\Delta t$, has to do with changes in the textual structure of the messages. At around this point, we see that the frequency of user mentions $p_@$ peaks, and the frequency of questions $p_?$ is about to drop dramatically, meaning that conversations become less and less interactive.

These findings, which are obtained by looking in the aggregate at averages over many conversations from many distinct channels, have also support at the individual level  (\textsc{rq}~5) -- for the majority of users, output $M_u$ follows an inverted U-shape curve as a function of information rate $V$, similarly to Figure~\ref{fig:overload}. To our knowledge, this is the first time evidence of information overload is observed at the individual level. We also provide evidence that the drop in participation cannot be explained in terms of heterogeneous room activity, since patterns similar to those of Figure~\ref{fig:overload} hold even when we break down the data by room activity (Figure~\ref{fig:overload_quartile}), nor by exogenous increases of $U$ --- for example due to a particularly good or entertaining move in the game --- because any increase in $U$ must correspond to an increase in $V$. This latter observation is due to a limitation of our data, which do not include the number of non-chatting viewers.

%

Our methodology is not perfect and some limitations must be acknowledged. The first has to do with text. Some of our textual features are based on English, like the list of discourse markers that we use to compute $p_d$. Of course there are plenty of non-English speakers on Twitch, and we do not filter out their messages in our data. Different discourse conventions from non-English speaking cultures may thus introduce bias in the results for $p_d$, and perhaps even of $p_?$. In part to mitigate for these assumptions, we apply content-agnostic features such as the compression ratio $\rho_c$. In the future we plan to use language detection techniques to select only a subset of languages, and use language-specific lists of discourse markers.

Another important limitation is that we do not have data to characterize the other main source of information in the Twitch UI --- the video feed itself. Arguably users must devote part of their cognitive load to process the video itself. It is reasonable to assume that chat activity is correlated to what happens in the feed, for example because viewers \emph{react} to what happens in the video.

Therefore, even though the presence of a second source of overload may skew our measurements about the onset of the overloaded phase (i.e., the precise location of $V^*$), the results about the change in nature of conversation (Figure~\ref{fig:features}) are unlikely to change much. While analyzing video brings significant technical challenges (e.g. synchronization, frame analysis, etc.), it would be interesting to incorporate it in at least a small case study.     

Does overload produce longer-lasting effects, besides what we already see in the Twitch chat? In line with previous literature on herding~\cite{Raafat2009} and social contagion~\cite{hodas2012visibility,gomez2014quantifying}, we speculate that overload has a strong effect on the production and dissemination of new memes. In future research we would like to investigate whether periods of overload are germane to the rise of popularity of new slang and expressions, and whether they are preferentially coined under conditions of overload.


In conclusion, we studied the dynamics of Twitch chat conversations. To our knowledge, this is the first time a large systematic sample of logs from the Twitch chat has been analyzed. We provide quantitative measurements for the onset of information overload at both the collective and individual level, and describe its effects on the overall structure and dynamic of the group. Our finding may inform designers of social media UIs. For example it could be beneficial to introduce an automatic detector of possible overload based on the rate of messages, or visual aids for users to cope with overload.



\section*{Competing interests}
  The authors declare that they have no competing interests.

\section*{Author's contributions}

AN, GLC, AF, and YA designed the research; AN and GLC performed experiments and analyzed the data and wrote the manuscript; all authors reviewed the manuscript.

\section*{Acknowledgements}

The authors would like to acknowledge Cyrus Hall, Drew Harry, Spencer Nelson, and Ruth Toner from Twitch for granting access to the chat logs data and for useful conversations. We would also like to thank Haewoon Kwak for insightful conversations. This work was partially supported by the Indiana University Network Science Institute. GLC acknowledges partial support from the Swiss National Science Foundation (fellowship PBTIP2\_142353).

\bibliographystyle{plain}
\bibliography{ref}

\end{document}